\documentstyle{article}
\input boxedeps.tex
\SetRokickiEPSFSpecial  
\HideDisplacementBoxes

\newcommand{\be}{\begin{equation}}
\newcommand{\eq}{\end{equation}}

\begin{document}
\mbox{SWAT/376}\hfil\\
\vspace{30mm}
\begin{center}IMPACT PARAMETER DEPENDENT QUARK DISTRIBUTIONS OF THE PION\\
\vspace{20mm}

Simon Dalley\\
\vspace{20mm}
Department of Physics, University of Wales Swansea \\
Singleton Park, Swansea SA2 8PP, United Kingdom\\
email: pydalley@swansea.ac.uk
\vspace{20mm}


\end{center}


We present first results for the impact parameter dependent
quark distribution in the pion obtained in transverse lattice gauge theory,
and discuss recent predictions from other models.



\newpage

\section{Introduction}
\label{intro}

First-principles QCD calculations of the structure of hadrons
are available from lattice gauge theory. In particular, the
low moments of traditional lightcone momentum fraction $x$ dependent
structure functions have been calculated \cite{lattice}.
Such moments are however consistent with a wide variety of
functional forms for the structure functions themselves. The latter
can be estimated directly in QCD-based models such as
truncated Dyson-Schwinger equations \cite{ds}, 
coarse transverse lattice gauge theory \cite{trans},
chiral quark models \cite{ruiz1,chiral} and instanton liquid \cite{inst}.
Typically, at low resolution scales these models disagree in detail
about the shape of quark distributions, although the differences
are mollified by evolution to higher scales, where experimental
data can be compared with.  
More general off-forward matrix elements, which are coherent
functions also of transverse parton momenta that interpolate
between traditional structure functions and form factors, have also
begun to be studied in these models. In particular, the traditional
structure functions may be generalised to ones depending also on the
transverse impact parameter $b$ of the struck quark, 
which retain a probabilitistic 
interpretation. Their experimental measurement
will provide  much 
stronger constraints on theory than hitherto available.

We present new results for impact parameter
dependent valence quark distribution functions of the pion, obtained via the
transverse lattice method. We find that the $x-b$ dependence does not
factorize, as is sometimes assumed in phenomenological models.
In particular, at large lightcone
momentum fraction $x$, the quark distribution is dominated by
small impact parameters $b < 2/3$ fm, while at small $x$ the opposite
is true. 
The result at large $x$  is shown to be consistent 
with a recent
construction of generalised parton distributions
using double distributions \cite{chris}, 
with chiral NJL quark models using the 
Pauli-Villars regulator \cite{vento}, and with a simple two-body
formula with sharp fall-off in transverse space.
Experimental measurement of the small-$b$ dependence 
could serve to distinguish
between these and other models, clarifying the behaviour of the 
integrated distribution  deduced previously from pion-nucleon
Drell-Yan scattering.
 
\section{transverse lattice results} 

The transverse lattice formulation of gauge theory 
represents the physical gluonic degrees of freedom by
gauge-covariant links of colour flux on a lattice transverse to a
null-plane of quantisation \cite{bard}. 
In a recent paper \cite{trans1}, van de Sande and the author applied 
Discrete Light Cone Quantisation (DLCQ)  \cite{dlcq} to this formulation
in order to estimate the lightcone wavefunctions of the pion.
Lightcone wavefunctions provide a fundamental starting
point for the investigation of hadronic observables represented
as matrix elements of currents.
The computation performed in ref.~\cite{trans1}
 was done to leading
order of the $1/N_{c}$ expansion and used a coarse transverse
lattice of spacing $a \sim 2/3$ fm. This generated a series of gauge-invariant
effective interactions, consistent with residual Lorentz symmetries,
the low energy truncation of which left a set of
couplings to be determined. The couplings were fixed by requiring
finiteness, optimisation of full Lorentz covariance, and 
phenomenological fits to masses $M_{\pi}$, $M_{\rho}$, 
the decay constant $f_{\pi}$, and string tension $\sqrt{\sigma}$.

We introduce lightcone 
coordinates $\{z^+=(z^0 + z^3)/\sqrt{2}, z^-=(z^0 - z^3)/\sqrt{2},
{\bf z}=(z^1,z^2)\}$. 
At lightcone time $z^+ = 0$, 
we will write a hadron state 
in the impact-parameter
representation
$|P^+, {\bf b}_{0} \rangle$. $P^+ = (P^0 + P^3)/\sqrt{2}$ is the total
hadron lightcone momentum, ${\bf b}_{0}$ is the transverse position
of the hadron, assumed spinless. 
In this representation, a generalised quark distribution
may be defined in lightcone gauge as the matrix  element \cite{diehl}
\be
{\cal I}(x,\xi,{\bf b}) = {1 \over \sqrt{1-\xi^2}}
\langle {P}_{\rm out}^{+}, {\bf b}^{\rm out}_{0} | 
\int {dz^- \over 4 \pi} e^{{\rm i} x P^+ z^-} \overline{q}(0, -z^-/2, {\bf b})
\gamma^+ q(0, z^-/2, {\bf b})
|{P}_{\rm in}^{+}, {\bf b}^{\rm in}_{0} \rangle \ , \label{skew}
\eq
where 
\be
\xi = {(P_{\rm in}-P_{\rm out})^+ \over (P_{\rm in}+P_{\rm out})^+}
\ \ , \ \ {\bf b}^{\rm out}_{0}=-{\xi \over 1-\xi} {\bf b} \ \ , \ \ 
{\bf b}^{\rm in}_{0}={\xi \over 1+\xi} {\bf b} \ ,
\eq
and $q$ $(\overline{q})$ is an (anti)quark field.
It can be made gauge-invariant by insertion of a light-like Wilson
line and is not to be confused with so-called $k_T$-unintegrated
parton distributions, which we do not discuss here. 
These are qualitatively different and have
problems with gauge invariance \cite{diehl}.
We concentrate on the case $\xi = 0$. This is the most 
physically intuitive since ${\cal I}$ is then  simply the probability
of a quark carrying fraction $x$ of the lightcone momentum $P^+$ when 
at transverse position
${\bf b}$ \cite{mat,pire}. 
It is also the most easily investigated in the context of DLCQ,
which discretizes lightcone momentum fractions. At non-zero $\xi$, one
would have to extrapolate the DLCQ cutoff in lightcone wavefunctions first,
before
using them to compute observables like Eq.~(\ref{skew}), rather than the easier
task of extrapolating the observables themselves.

As well as the DLCQ regulator, which we extrapolate in observables,
the transverse lattice theory discretizes the transverse co-ordinate
${\bf z}$ on a square lattice of spacing $a \sim 2/3 {\rm fm}$, held fixed,
 and
decomposes the hadron state into its parton constituents. As well
as the quarks $q$ localised at transverse sites, other partons
consist of gauge potentials $A_{\pm}$, which are eliminated as
dynamical degrees of freedom by lightcone gauge choice $A_{-}=0$,
and $N_{c}$x$N_{c}$ matrix link-fields 
$M_{r}(z^-,{\bf z})$, which carry color flux
from site ${\bf z}$ to ${\bf z}+ a\hat{\bf r}$, where $\hat{\bf r}$ is a unit
vector in direction $r \in \{1,2\}$. 
$M_{r}^{\dagger}(z^-,{\bf z})$ is the oppositely oriented link and we
label the transverse orientation of link fields
by indices $\lambda_j \in \{\pm 1, \pm 2\}$.
By convention, we will associate the 
center of the link, ${\bf z}+ 0.5a\hat{\bf r}$, as the transverse
position of the link-field
parton. In the large-$N_c$ limit, a meson state is then written as
a superposition Fock states consisting of a $q$-$\overline{q}$ pair 
joined by various configurations of a string of link fields 
(to ensure residual transverse gauge invariance in the lightcone gauge).

In ref.~\cite{trans1},
 the partonic decomposition was done for a hadron state
$|P^+, {\bf P} \rangle$
fully in momentum space. In particular, the wavefunctions
$\psi_n(x_{i}, h, h^{\prime}, \lambda_{j})$ 
were computed\footnote{The raw DLCQ data for wavefunctions
$\psi_{n}$ found in ref.\cite{trans1} are available
from the internet URL  http://www.geneva.edu/~bvds/four/}
 for the pion in the case
\begin{eqnarray}
|P^+, {\bf P} & = & {\bf 0} \rangle = \sum_{n=2}^{\infty} \int [dx]_{n}
\sum_{{\bf z}_1, {\bf z}_n, h, h^{\prime}}
\left[\sum_{\lambda_j}\right]_{n}
\psi_n(x_{i}, h, h^{\prime}, \lambda_{j})
|(x_{1}, h, {\bf z}_{1}); 
(x_2, \lambda_1, {\bf z}_{2}) ); \cdots  \nonumber  \\
&& \cdots ; (x_{n-1}, \lambda_{n-2}, {\bf z}_{n-1}); 
(x_{n}, h^{\prime}, {\bf z}_{n})  \rangle \ , 
\label{decom}
\end{eqnarray}
where $h, h^{\prime}$ denote quark and anti-quark helicities respectively,
$x_i$ is the fraction of $P^+$ carried by the $i^{\rm th}$ parton,
\be 
\int [dx]_{n}  =  \int dx_1 \cdots dx_n \delta\left( \sum_{i=1}^{n}
 x_i - 1\right) 
\eq
and $\left[\sum_{\lambda_j}\right]_{n}$ indicates that the sum over
orientations of links must form an unbroken chain on the transverse lattice
between quark and anti-quark.
Note that once the transverse positions ${\bf z}_1$ and ${\bf z}_{n}$
of the quark and antiquark are specified, it is sufficient to just 
enumerate the sequence of link orientations $\lambda_j$ rather than their
actual postions ${\bf z}_{j}$.

A set of hadron states boosted to general transverse momentum ${\bf P}$
can be obtained by applying
the Poincar\'e generators ${\bf M}^{+} = (M^{+1}, M^{+2})$. 
This gives for each parton Fock state
\begin{eqnarray} 
&&{\rm exp}\left[ -{\rm i} {\bf M}^{+}. {\bf P}/P^+ \right]
|(x_{1}, h, {\bf z}_{1}); 
(x_2, \lambda_1, {\bf z}_{2}) ); \cdots  ;
(x_{n-1}, \lambda_{n-2}, {\bf z}_{n-1}); 
(x_{n}, h^{\prime}, {\bf z}_{n})  \rangle \nonumber \\
&& = 
{\rm exp}\left[ {\rm i} {\bf P}. {\bf c} \right]
|(x_{1}, h, {\bf z}_{1}); 
(x_2, \lambda_1, {\bf z}_{2}) ); \cdots  ;
(x_{n-1}, \lambda_{n-2}, {\bf z}_{n-1}); 
(x_{n}, h^{\prime}, {\bf z}_{n})  \rangle \ , \label{boost}
\end{eqnarray}
\be
{\bf c} = (c^1 , c^2) = \sum_{i=1}^{n} x_i {\bf z}_{i} \ .
\eq
Before investigating the interplay of transverse and longitudinal
structure, we check the special case of the electromagnetic form factor
\be
\langle P_{\rm in}^{+}, {\bf P}_{\rm in} |
j^{\mu}_{\rm em}
|P_{\rm out}^{+}, {\bf P}_{\rm out} \rangle = F(Q^2) (P_{\rm in} + 
P_{\rm out})^{\mu} \label{form}
\eq
where $Q = P_{\rm out} - P_{\rm in}$.
Using the decomposition (\ref{decom}), boosts (\ref{boost}), and wavefunctions
$\psi_n$,
$F(Q^2=-t)$ may be extracted
most simply from the $\mu=+$ component of the matrix element (\ref{form})
in the case of purely transverse momentum transfer $Q$ (Drell-Yan frame)
\be
F(Q^2) = \sum_{n=2}^{\infty}
\int [dx]_{n}\sum_{{\bf z}_1, {\bf z}_n, h, h^{\prime}}
\left[\sum_{\lambda_j}\right]_{n} {\rm exp}
\left[ {\rm i} {\bf Q}. {\bf c} \right]
|\psi_n(x_{i}, h, h^{\prime}, \lambda_{j})|^2 \ . \label{form1}
\eq
Figure 1 shows that calculations 
of the $\psi_n$ truncated at $n=3$ \cite{trans2} and $n=5$ \cite{trans1}
give a form factor converging  close to the
experimental one (it should be noted that the latter suffers 
from uncertainties due to the extrapolation
to the pion pole). A cutoff on $n$ cuts off the maximum quark-antiquark
separation.
With the $n=5$ truncation, the maximum quark antiquark
separation is $3a \sim 2$ fm, which explains why the calculated charge radius
$r_{\pi}= \sqrt{6dF(-t)/dt}|_{t=0} \approx 0.59 $~fm  is slightly less than
the experimental value $0.663$~fm. 

The traditional quark distribution function 
$V(x) = \sum_{\bf b} {\cal I}(x,0,{\bf b})$ was already given in
ref.\cite{trans1} 
\be
V(x) = {(1-x)^{0.33} \over x^{0.7}}\left\{ 0.33 -1.1\sqrt{x} + 2x \right\}
\eq
and 
is shown in Figure 2(a) at a transverse resolution scale of 
$0.5 {\rm GeV}$, set roughly
by the lattice spacing $a$.
As manifestly
required of many-body gauge theory lightcone wavefunctions, 
it satisfies the usual normalisation and sum rules, rising Regge behaviour 
at small $x$ and vanishing at $x \to 1$. The latter two properties are finite
energy conditions \cite{anton}.

To investigate simultaneously the transverse and longitudinal structure
we can use the impact parameter dependent parton distributions
${\cal I}(x,0,{\bf b})$.
We must first translate results to hadron impact 
parameter space in the transverse directions. 
Integrating ${\bf P}$ over the Brillouin zone, we obtain a hadron
state localised at transverse position ${\bf 0}$
\begin{eqnarray}
|P^+, {\bf b}_{0} ={\bf 0} \rangle> & = & 
\sum_{n=2}^{\infty} \int [dx]_{n}
\sum_{{\bf z}_1, {\bf z}_n, h, h^{\prime}}
\left[\sum_{\lambda_j}\right]_{n}
{a^2 \over \pi^2} {\sin{c^{1}\pi/a} \over c^1}{\sin{c^{2}\pi/a} \over c^2}
\psi_n(x_{i}, h, h^{\prime}, \lambda_{j}) \nonumber \\
&& \times |(x_{1}, h, {\bf z}_{1}); 
(x_2, \lambda_1, {\bf z}_{2}) ); \cdots  ;
(x_{n-1}, \lambda_{n-2}, {\bf z}_{n-1}); 
(x_{n}, h^{\prime}, {\bf z}_{n}) \ .  \rangle
\end{eqnarray}
Hence
\be
{\cal I}(x,0,{\bf b}) = \sum_{n=2}^{\infty}
\int [dx]_{n}\sum_{{\bf z}_1, {\bf z}_n, h, h^{\prime}}\left[\sum_{\lambda_j}\right]_{n} \delta(x_1 -x) \delta({\bf z}_1 -
{\bf b})
\left[{a^2 \over \pi^2} {\sin{c^{1}\pi/a} \over c_1}
{\sin{c^{2}\pi/a} \over c_2}\right]^2
|\psi_n(x_{i}, h, h^{\prime}, \lambda_{j})|^2 \ . 
\eq
Due to the lattice cutoff, this should strictly be interpreted
as the probability of finding the quark within one lattice spacing
of impact parameter ${\bf b}$. We show results for the first few
values of $b=|{\bf b}|$ sampled along lattice axes in fig.2(b)-(e). 
Fig. 2(b) is the contribution from ${\bf b} = (0,0)$, Fig. 2(c)
is the sum of contributions from ${\bf b} = (a,0),(0,a),(-a,0),(0,-a)$,
etc..
The equations of these curves are fits to extrapolated DLCQ data
\begin{eqnarray}
b=0 & : & 0.18(1-4.3\sqrt{x}+11x)x^{-0.16}(1-x)^{0.37} \nonumber \\
b=a & : & 0.11(1-1.9\sqrt{x}+5.5x)x^{-0.78}(1-x)^{2.0}  \nonumber \\
b=2a & : & 0.04(1-4.2\sqrt{x}+6.2x)x^{-0.76}(1-x)^{1.94} \nonumber \\
b=3a & : & 0.005(1-4.0\sqrt{x}+8.2x)x^{-0.66}(1-x)^{1.2} \ .
\label{fits}
\end{eqnarray}
The figure clearly
shows that the valence region $x > 0.5$ is dominated by quarks
with impact parameter less than the charge radius $\sim 2/3$~fm.
It also suggests that the rapid rise at small $x$ is dominated
by larger impact parameters $b > 2/3$~fm. 

The above results have an intuitive explanation. The transverse lattice
formulation is non-perturbatively gauge invariant. As a result of
local gauge invariance in the transverse direction --- an axial gauge fixing is
applied in other directions ---  quarks and antiquarks must be joined by
a  string of flux-carrying link variables. Since these link variables
are themselves also dynamical parton degrees of freedom carrying a
portion of the hadron $P^+$, they cost energy. At large quark $x$, all other 
partons are forced to carry small $x$ and it is favourable to have the
minimum number of them. The suppression of link partons and gauge
invariance ensures widely separated quarks are suppressed in this
region. We observe that this behaviour sets in quite quickly after $x= 0.5$.
The rise at small $x$ is due to higher Fock states \cite{anton} containing
more link partons, which naturally allows more widely separated quarks
to contribute.

The analytic dependence on ${\bf b}$ is difficult to characterize
due to that fact that only a coarse sampling is possible and
this represents averages over unit lattice cells centred at ${\bf b}$. 
However, we found
that the following transverse-continuum two-body form, for quarks at transverse
positions ${\bf z}_1, {\bf z}_2$,
 reproduces the qualitative features of
the distributions in Fig.~2 for $x > 0.2$, 
\be
{\cal I}(x,0,{\bf b}) = { 8V(x) \over 
3\pi\left( 1+ {|{\bf z}_1|^2 + |{\bf z}_2|^2 \over 3a^2}\right)^3} 
\left(1+{x^2 \over (1-x)^2}\right) \ . 
\label{analytic}
\eq
We identify $b= |{\bf z}_1|$ and the positions are constrained 
by the fixed hadron transverse `center-of-momentum'
\be
{\bf c} =  x {\bf z}_1 + (1-x) {\bf z}_2 = {\bf 0} \ . 
\eq
The other $x$-dependent factors in (\ref{analytic}) have been chosen simply 
so that
$\int d^2 {\bf b} \ {\cal I}(x,0,{\bf b}) = V(x)$. It does not completely
fit the small-$x$ behaviour, but a two-body formula is not expected
to work well in this region. Eq.(\ref{analytic}) implies a
sharp fall-off of the hadron wavefunction in transverse space at a 
particular ($x$-dependent) radius. 

\section{Discussion}
Generalised quark distributions of the
pion have recently been
analysed in other models. Mukherjee {\em et al.} \cite{chris}
construct generalised
parton distributions from 
double distributions, which satisfy general reduction, spectral, and
polynomiality conditions. (Their construction has been 
criticised on grounds of positivity \cite{gerry}, 
but this does not affect the
discussion here.) 
We could not obtain the impact parameter dependent
distribution in closed form from their formulas, but one may 
deduce its transform
\begin{eqnarray}
\int d^2{\bf b} \ {\rm exp}\left[ {\rm i} {\bf b}.{\bf q} \right]
{\cal I}(x, 0 , {\bf b}) & \propto 
& { (1-x) \over \sqrt{x}} \lambda^3 A(\lambda) \label{double}
\\
\lambda & = & 
-{16 m^2 x \over t s^2 (1-x)}\left( 1 + {s^2 (x-0.5)^2 \over x(1-x)}
\right)\\
A(\lambda) & = & {\lambda -2\over \lambda(1+\lambda)^2}
-{4+\lambda \over (1+\lambda)^{5/2}}
\log\left[{1+\sqrt{1+\lambda} \over  \sqrt{\lambda}}\right]
\end{eqnarray}
where $t=-{\bf q}^2 < 0$, while $m=0.46$~GeV and $s=0.81$ are parameters
they adjusted for a fit to the pion form factor. 
As $x \to 1$, the expression (\ref{double}) becomes $t$-independent, i.e.
dominated by the contribution from ${\bf b}={\bf 0}$. This is consistent
with the general behaviour expected as $x \to 1$ \cite{mat} and the
transverse lattice result.
The result is independent of the tunable parameters in the
construction and any possible modifications at small $x$ necessary
to account for positivity \cite{chris}.

Generalised parton distributions have 
been calculated for the pion in chiral NJL models using the Pauli-Villars
regulator \cite{vento}. This respects symmetries of the model and 
accurately fits a wide range of 
experimental data. It has the striking result that, 
at the cut-off scale
of the model $\sim 0.3-0.4$~GeV, 
the quark distribution function 
$V(x)$ is constant in the chiral limit \cite{ruiz1}. This is, in fact, 
not far from the transverse lattice result, which is associated with a 
slightly higher resolution scale.
The authors of ref.\cite{vento} find that the transformed impact parameter
dependent distribution (\ref{double})
rises at large $x$ for large $-t$, consistent
with the transverse lattice result. It is significant that the results
in the NJL model and transverse lattice are so similar. The NJL model
treats chiral symmetry transparently, yet is an effective quark theory
containing no explicit gluonic degrees of freedom. The transverse lattice
construction generically (explicitly) breaks chiral symmetry, 
which one tries to minimze by tuning of couplings, yet incorporates
explicitly the confining effects of multiple gluonic degrees of freedom.

A new approach using a spectral quark model has also been used 
to obtain the same constant valence lightcone wavefunction of the pion in 
ref.\cite{ruiz2}. It would be interesting to know the result this
model gives for the impact parameter dependence.
Results from truncated Dyson-Schwinger
equations \cite{ds2}, the Instanton Liquid \cite{tomio}, and other
models \cite{models} 
are also available for the traditional distribution that is integrated 
over $b$. The detailed shapes, even for the integrated distributions
$V(x)$, generally differ amongst all the models considered in this paper,
although this could in part be due to different renormalisation scales
and schemes.

Existing data on $V(x)$  from pion-nucleon Drell-Yan
scattering \cite{conway} largely agrees with the 
NJL model (Pauli-Villars regulated) 
and the transverse lattice, but shows some differences in the valence
region (large $x$) of other models, notably the otherwise 
reliable Dyson-Schwinger
equations \cite{ds2} and the Instanton Liquid \cite{tomio}. 
Experimental measurement of impact parameter dependent distributions
would throw light on this puzzle.
In principle they can be found from wide-angle real Compton scattering
off pions. Data collected above a certain large transverse momentum transfer
$\sqrt{-t}$ could be compared, as a function of $x$,  with the full
valence quark distribution function $V(x)$ of the pion, which corresponds to
$t=0$. The transverse
lattice result would predict that they are the same at large $x$.

\vspace{10mm}
\noindent {Acknowledgements}:
This work was supported by PPARC grant PPA/G/O/2000/00448. I thank
M. Burkardt and W. Broniowski for helpful comments on the manuscript.

\begin{figure}
\centering
\BoxedEPSF{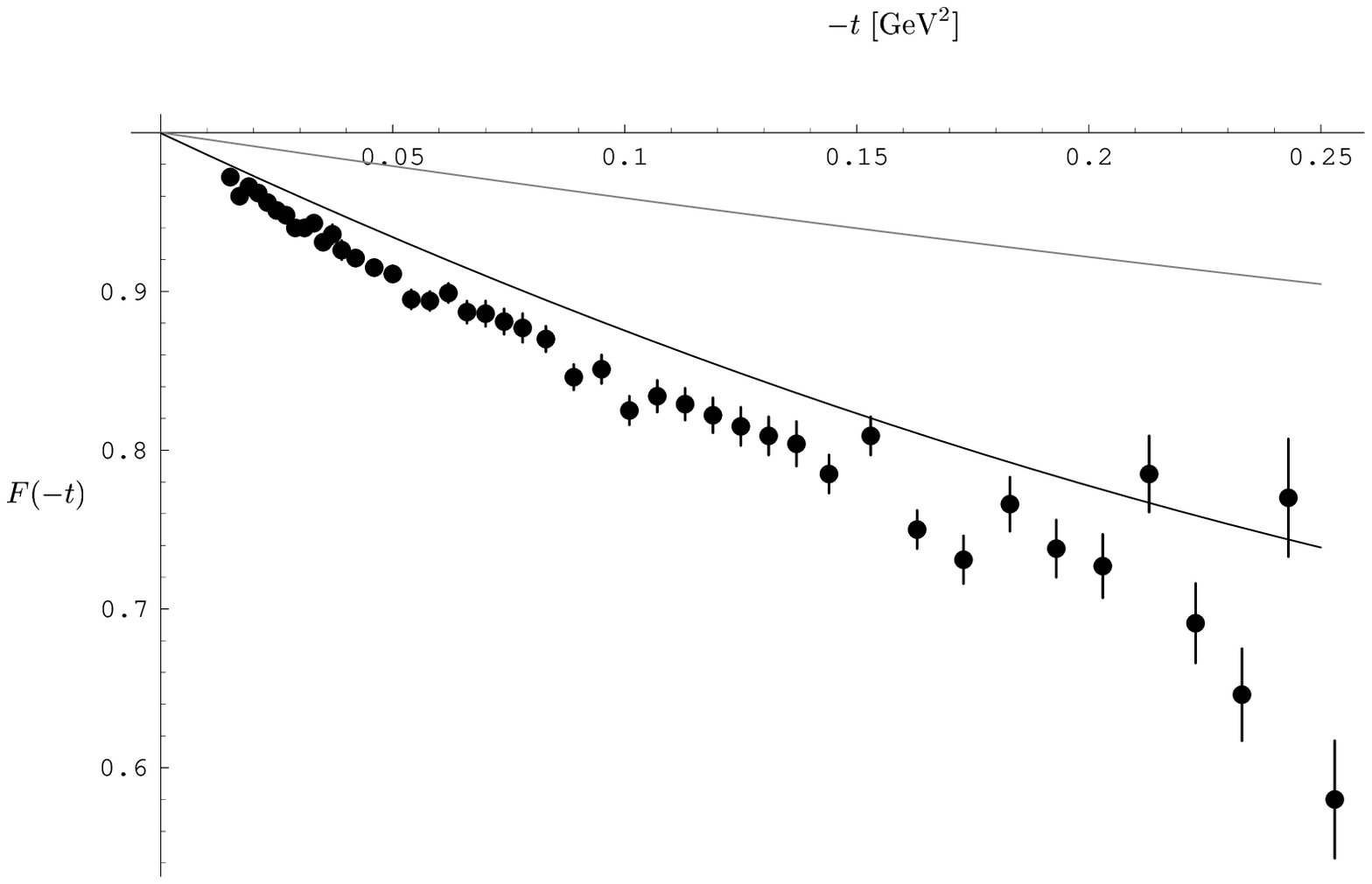 scaled 800}\\
\vspace{-50mm}
\caption{Pion form factor $F(-t)$. Solid curves are the transverse
lattice result from eq.(\ref{form1}), gray curve for wavefunctions calculated
within a truncation at $n=3$, black curve for $n=5$. The experimental data
points are from ref.\cite{exp}.
\label{fig1}}
\end{figure}

\begin{figure}
\centering
\BoxedEPSF{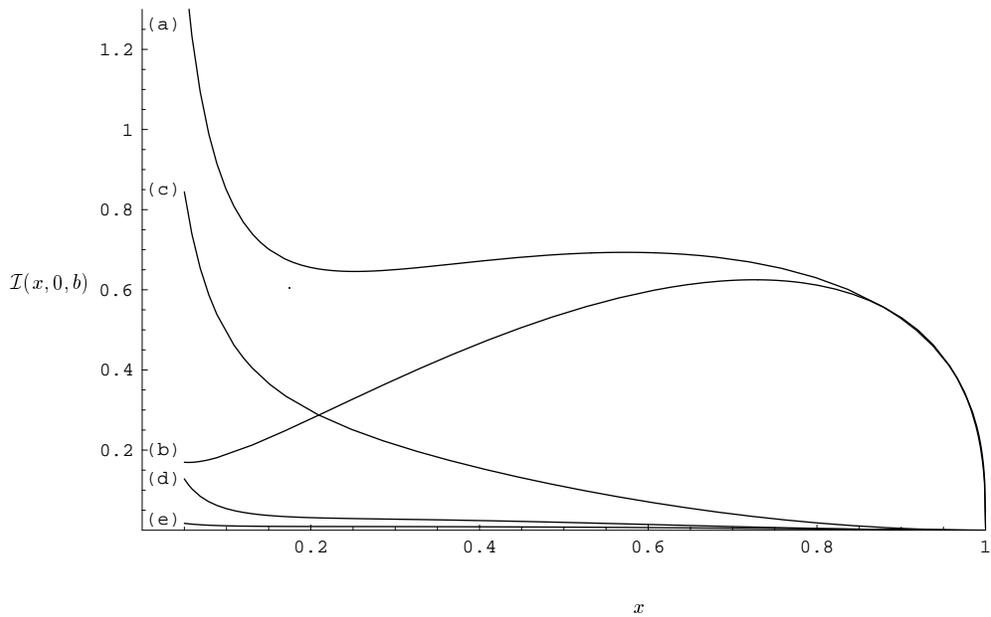 scaled 800}\\
\vspace{-50mm}
\caption{Impact parameter dependent valence quark distributions of the
pion. (a) The full distribution $V(x)$, which  sums 
over all impact parameters ${\bf b}$. (b)-(e) The sum of contributions 
at impact parameters ${\bf b}$ along  lattice axes such that 
$b= | {\bf b}| =0,a,2a,3a$ respectively. 
\label{fig2}}
\end{figure}

\end{document}